\documentclass[10pt,aps,prd,twocolumn,notitlepage,bibnotes,longbibliography,
floatfix,showpacs,citeautoscript,superscriptaddress]{revtex4-1} 

\usepackage[english]{babel}	%%% LaTeX for English
\usepackage{amsmath}		%%% Mathematics AMS package 
\usepackage{amssymb}		%%% More math symbols
\usepackage{bbm}			%%% Font for ensembes
\usepackage{graphicx}		%%% Include graphics
\usepackage{graphics}		%%% Include graphics
\usepackage{color}			%%% Color options for eps
\usepackage{xcolor}			%%% Color options for eps
\DeclareGraphicsExtensions{.pdf,.jpg,.jpeg,.png,.eps}	%%% Extensions for pdflatex
\usepackage[colorlinks=true, pdfstartview=FitV,linkcolor=blue, 
			citecolor=blue, urlcolor=blue]{hyperref}
\newcommand{\Equation}[2]{  \begin{equation}\label{#1}#2\end{equation} }
\newcommand{\Align}[2]{\begin{align}\label{#1}#2\end{align}}
\newcommand{\SubAlign}[2]
{\begin{subequations}\label{#1}\begin{align}#2\end{align}\end{subequations}}
\newcommand{\Vector}[1]{\left( \begin{array}{c}#1\end{array}\right)}
\newcommand{\bs}{\boldsymbol}
\newcommand{\Figref}[1]{Fig.~\ref{#1}}
\newcommand{\Eqref}[1]{\eqref{#1}}
\newcommand{\Exp}[1]{\text{e}^{#1}}
\renewcommand\Re{\mathrm{Re}}
\renewcommand\Im{\mathrm{Im}}
\newcommand{\Grad}{{\bs \nabla}}

\newcommand{\Curl}{{\bs \nabla}\!\times\!}
\newcommand{\x}{\bs x}

\newcommand{\F}{\mathcal{F}}
\newcommand{\D}{{\bs D}}
\newcommand{\A}{{\bs A}}
\newcommand{\B}{{\bs B}}
\newcommand{\J}{{\bs J}}
\renewcommand{\j}{{\bs j}}
\newcommand{\vv}{{\bs v}}
\newcommand{\p}{{\bs p}}
\newcommand{\tx}{\tilde{\bs x}}
\newcommand{\tGrad}{\tilde{\bs \nabla}}

\newcommand{\tCurl}{\tilde{\bs \nabla}\!\times\!}
\newcommand{\trho}{\tilde{\rho}}
\newcommand{\tz}{\tilde{z}}

\newcommand{\beqn}{\begin{eqnarray}}
\newcommand{\eeqn}{\end{eqnarray}}
\newcommand{\eq}[1]{(\ref{#1})}

\begin{document}
%%%%%%%%%%%%%%%%%%%%%%%%%%%%%%%%%%%%%%%%%%%%%%%%%%%%%%%%%%%%%%%%%%%%%
%%%%%%%%%%%%%%%%%%%%%%%%%%%%%%%%%%%%%%%%%%%%%%%%%%%%%%%%%%%%%%%%%%%%%
%%%% Title informations and authors
\title{Vortices with magnetic field inversion in noncentrosymmetric superconductors}

\author{Julien Garaud}
\email{garaud.phys@gmail.com}
\affiliation{Institut Denis Poisson CNRS/UMR 7013, 
			 Universit\'e de Tours, 37200 France}
\author{Maxim N. Chernodub}
\email{maxim.chernodub@idpoisson.fr}
\affiliation{Institut Denis Poisson CNRS/UMR 7013, 
			 Universit\'e de Tours, 37200 France}
\affiliation{Pacific Quantum Center, 
			 Far Eastern Federal University, Sukhanova 8, Vladivostok, 690950, Russia}
\author{Dmitri E. Kharzeev}
\email{dmitri.kharzeev@stonybrook.edu}
\affiliation{Department of Physics and Astronomy, 
			 Stony Brook University, New York 11794-3800, USA}
\affiliation{Department of Physics and RIKEN-BNL Research Center,\\ 
			 Brookhaven National Laboratory, Upton, New York 11973, USA}
\affiliation{Le Studium, Loire Valley Institute for Advanced Studies, 
			 Tours and Orl\'eans, France}

\date{\today}

%%%%%%%%%%%%%%%%%%%%%%%%%%%%%%%%%%%%%%%%%%%%%%%%%%%%%%%%%%%%%%%%%%%%%
\begin{abstract}
Superconducting materials with noncentrosymmetric lattices lacking space inversion 
symmetry exhibit a variety of interesting parity-breaking phenomena, including the
magneto-electric effect, spin-polarized currents, helical states, and the unusual 
Josephson effect. We demonstrate, within a Ginzburg-Landau framework describing 
noncentrosymmetric superconductors with $O$ point group symmetry, that vortices can 
exhibit an inversion of the magnetic field at a certain distance from the vortex core. 
In stark contrast to conventional superconducting vortices, the magnetic-field reversal 
in the parity-broken superconductor leads to non-monotonic intervortex forces, and, 
as a consequence, to the exotic properties of the vortex matter such as the formation of 
vortex bound states, vortex clusters, and the appearance of metastable vortex/anti-vortex 
bound states.
\end{abstract}
%%%%%%%%%%%%%%%%%%%%%%%%%%%%%%%%%%%%%%%%%%%%%%%%%%%%%%%%%%%%%%%%%%%%%

\maketitle

%%%%%%%%%%%%%%%%%%%%%%%%%%%%%%%%%%%%%%%%%%%%%%%%%%%%%%%%%%%%%%%%%%%%%
\section{Introduction}\label{Introduction}

noncentrosymmetric superconductors are superconducting materials whose crystal 
structure is not symmetric under the spatial inversion. These parity-breaking 
materials have attracted much theoretical \cite{Bulaevskii.Guseinov.ea:76,
Levitov.Nazarov.ea:85,Mineev.Samokhin:94,Edelstein:96,Agterberg:03,Samokhin:04} 
and experimental \cite{Bauer.Hilscher.ea:04,Samokhin.Zijlstra.ea:04,Yuan.Agterberg.ea:06, 
Cameron.Yerin.ea:19,Khasanov.Gupta.ea:20} interest, as they make it possible to 
investigate spontaneous breaking of a continuous symmetry in a parity-violating 
medium (for recent reviews, see \cite{Bauer.Sigrist,Yip:14,Smidman.Salamon.ea:17}). 
The parity-breaking nature of the superconducting order parameter 
\cite{Samokhin.Zijlstra.ea:04,Yuan.Agterberg.ea:06} in noncentrosymmetric 
superconductors leads to various unusual magneto-electric phenomena due to the mixing 
of singlet and triplet components of the superconducting condensate, correlations 
between supercurrents and spin polarization, to the existence of helical states, 
and to unusual structure of vortex lattices.

Moreover, parity breaking in noncentrosymmetric superconductors also results 
in an unconventional Josephson effect, where the junction features a phase-shifted 
relation for the Josephson current~\cite{Buzdin:08,Konschelle.Buzdin:09}. 
Unconventional Josephson junctions consisting of two noncentrosymmetric 
superconductors linked by a uniaxial ferromagnet were recently proposed as 
the element of a qubit that avoids the use of an offset magnetic flux, enabling 
a simpler and more robust architecture \cite{Chernodub.Garaud.ea:19}.

\begin{figure}[!htb]
\hbox to \linewidth{ \hss
\includegraphics[width=0.95\linewidth]{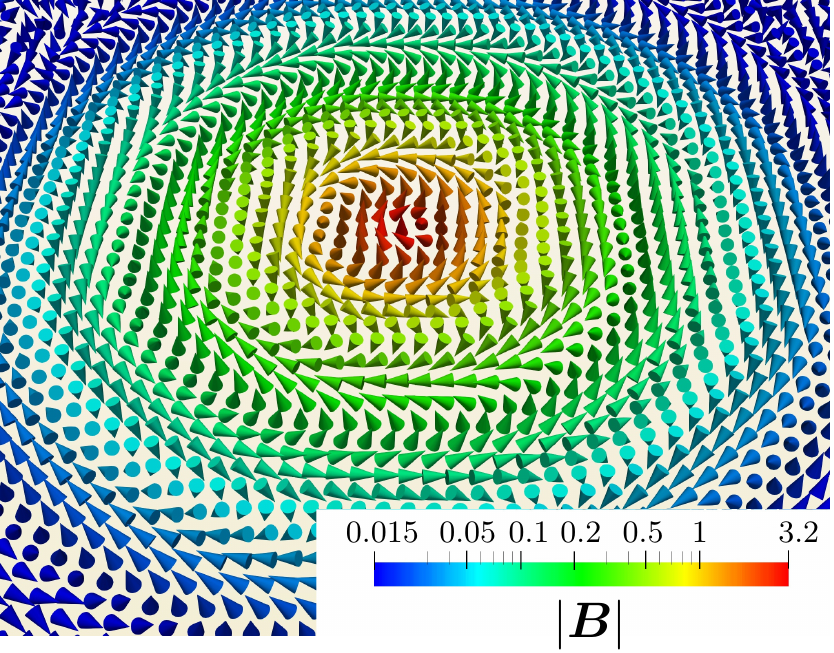}
\hss}
\caption{ 
Inversion patterns of the magnetic field $\B$ of a vortex in a noncentrosymmetric 
superconductor. The magnetic field forms helicoidal patterns around a straight 
static vortex. As the distance from the vortex core increases, the longitudinal 
(parallel to the vortex core) component of the magnetic field may change its sign. 
The magnetic field may exhibit several sign reversals in the normal plane. In the 
picture, which is a result of a numerical simulation of the Ginzburg-Landau theory, 
the colors encode the amplitude of the magnetic field $\B$, in a normal plane with 
respect to the vortex line while the arrows show the orientation of the field. 
}
\label{Fig:Field-inversion}
\end{figure}

In the macroscopic description of such superconducting states, the lack of inversion 
symmetry yields new terms in the Ginzburg-Landau free energy termed `Lifshitz invariants'. 
These terms directly couple the magnetic field with an electric current and thus lead 
to a variety of new effects that are absent in conventional superconductors. The 
explicit form of the allowed Lifshitz invariant depends on the point symmetry group 
of the underlying crystal structure.

In this paper, we consider a particular class of noncentrosymmetric superconductors 
that break the discrete group of parity reversals and, at the same time, are invariant 
under spatial rotations. The corresponding Lifshitz invariant featuring these symmetries 
is described by the simple, parity-violating isotropic term, 
$\propto\Im(\psi^*\D\psi)\cdot\B$. This particular structure describes noncentrosymmetric 
superconductors with $O$ point group symmetry such as Li$_2$Pt$_3$B 
\cite{Badica.Kondo.ea:05,Yuan.Agterberg.ea:06}, Mo$_3$Al$_2$C \cite{Karki.Xiong.ea:10,
Bauer.Rogl.ea:10}, and PtSbS \cite{Mizutani.Okamoto.ea:19}.

Vortex states in cubic noncentrosymmetric superconductors feature a transverse 
magnetic field, in addition to the ordinary longitudinal field. Consequently, 
they also carry a longitudinal current on top of the usual transverse screening 
currents \cite{Lu.Yip:08,Lu.Yip:08a,Kashyap.Agterberg:13}.
Therefore, as illustrated in \Figref{Fig:Field-inversion}, both the superconducting 
current and the magnetic field form a helical-like structure that winds around the 
vortex core (for additional material illustrating the helical spatial structure of 
the magnetic streamlines, see Appendix \ref{Sec:Helicity}, and supplemental animations 
\cite{
[{See Supplemental video material: }]
[{}] Supplemental-arxiv} 
described in Appendix \ref{Sec:Movies}). 
Previous theoretical papers studied vortices in the perturbative regime where the 
coupling to the Lifshitz invariant is small, either in the London limit (with a large 
Ginzburg-Landau parameter) \cite{Lu.Yip:08,Lu.Yip:08a}, or beyond it 
\cite{Kashyap.Agterberg:13}. 
For currently known noncentrosymmetric materials, these approximations are valid 
since the magnitude of the Lifshitz invariants, which can be estimated in a 
weak-coupling approximation, is typically small. We propose here a general study of 
vortices, for all possible values of the Lifshitz invariant coupling, both in the 
London limit and beyond.

We demonstrate that vortices may feature an inversion of the magnetic field at a
distance of about 4$\lambda_L$ from the vortex center. Moreover, for rather high 
values of the coupling $\gamma$, alternating reversals may occur several times, at 
different distances from the vortex core. Such an inversion of the magnetic field 
is illustrated, in \Figref{Fig:Field-inversion}. The reversal of the magnetic field, 
which is in stark contrast to conventional superconducting vortices, becomes 
increasingly important for larger couplings of the Lifshitz invariant term.
This property of field inversion is responsible for other unusual behaviors, also 
absent in conventional type-2 superconductors. Indeed, we show that it leads to the 
formation of vortex bound states, vortex clusters, and meta-stable pairs of vortex 
and anti-vortex. These phenomena should have numerous physical consequences on the 
response of noncentrosymmetric superconductors to an external magnetic field.

The paper is organized as follows. In Sec.~\ref{Sec:Theory}, we introduce the 
phenomenological Ginzburg-Landau theory that describes the superconducting state 
of a noncentrosymmetric material with $O$ point group symmetry.  Next, in 
Sec.~\ref{Sec:Vortices} we investigate the properties of single vortices both in 
the London limit and beyond it. We also demonstrate that the parity-breaking 
superconductors can feature an inversion of the longitudinal magnetic field. 
This observation suggests that the intervortex interaction in parity-odd 
superconductors might be much richer than that for a conventional superconductor. 
Hence we derive analytically the intervortex interaction energy in the London limit 
in Sec.~\ref{Sec:Interactions}. We show that the interaction potential depends 
non-monotonically on the intervortex distance, which leads to the existence of 
vortex bound states. Using numerical minimization of the Ginzburg-Landau free 
energy, we further observe that such bound states persist beyond the London limit. 
Our conclusions and discussion of further prospects are given in the last section.

%%%%%%%%%%%%%%%%%%%%%%%%%%%%%%%%%%%%%%%%%%%%%%%%%%%%%%%%%%%%%%%%%%%%%
%%%%%%%%%%%%%%%%%%%%%%%%%%%%%%%%%%%%%%%%%%%%%%%%%%%%%%%%%%%%%%%%%%%%%
%%%%%%%%%%%%%%%%%%%%%%%%%%%%%%%%%%%%%%%%%%%%%%%%%%%%%%%%%%%%%%%%%%%%%
\section{Theoretical framework}
\label{Sec:Theory}

We consider noncentrosymmetric superconductors with the crystal structure 
possessing the $O$ point group symmetry. Such materials are described
by the Ginzburg-Landau 
free energy $F=\int d^3x \, \F$ with the free-energy density given by 
(see e.g. \cite{Bauer.Sigrist,Agterberg:12}):
\Equation{Eq:FreeEnergy}{
 \F = \frac{\B^2}{8\pi}+k|\D\psi|^2+\gamma\j\cdot\B 
 +\frac{\beta}{2}(|\psi|^2-\psi_0^2)^2 \,,
}	 
where $\j=2e\,\Im\left(\psi^*\D\psi\right)$; we use $\hbar {=} c {=} 1$. 
Here, the single component order parameter $\psi=|\psi|\Exp{i\varphi}$ is a 
complex scalar field that is coupled to the vector potential $\A$ of the magnetic 
field $\B=\Curl\A$ through the gauge derivative $\D\equiv\Grad-ie\A$, where $e$ 
is a gauge coupling.
The explicit breaking of the inversion symmetry is accounted for by the Lifshitz invariant 
term with the prefactor $\gamma$, which directly couples the magnetic field $\B$ and 
$\j=2e|\psi|^2(\Grad\varphi-e\A)$. In the absence of parity breaking, when $\gamma=0$, 
the vector $\j$ matches the usual superconducting current. The parameter $\gamma$, 
which controls the strength of the parity breaking, can be chosen to be positive without 
loss of generality. The other coupling constants $k$ and $\beta$ describe, respectively, 
the magnitude of the kinetic and potential terms in the free energy \Eqref{Eq:FreeEnergy}.

The variation of the free energy \Eqref{Eq:FreeEnergy} with respect to the scalar 
field $\psi^*$ yields the Ginzburg-Landau equation for the superconducting condensate,
\Equation{Eq:EOM:GL}{
\big[k\D+2ie\gamma\B\big]\cdot\D\psi=\beta(|\psi|^2-\psi_0^2)\psi\,,
}
while the variation of the free energy with respect to the gauge potential $\A$ 
gives the Amp\`ere-Maxwell equation:
\Equation{Eq:EOM:AM}{
\Curl\Big(\frac{\B}{4\pi}+\gamma\j\Big)=k\j+2\gamma e^2|\psi|^2\B\equiv\J	\,.
}  
The supercurrent $\J$, is defined via the variation of the free energy 
\Eqref{Eq:FreeEnergy} with respect to the vector potential: $\J=\delta\F/\delta\A$. 
Nonzero parity-breaking coupling $\gamma$ gives an additional contribution from 
the Lifshitz term, which is proportional to $\B$ \cite{Agterberg:12} (see also remark 
\footnote{
Note that $\j$ matches the supercurrent $\J$ only when $\gamma=0$. Nonzero 
parity-breaking coupling $\gamma$ gives an additional contribution from 
the Lifshitz term to the supercurrent $\J$. We thus denote $\j$ to be a current, 
keeping in mind that the superconducting Meissner current is 
$\J=k\j+2\gamma e^2|\psi|^2\B$.}).
The physical length scales of the theory are, respectively, the coherence length $\xi$ 
and the London penetration depth $\lambda_L$,
\Equation{Eq:LengthScales}{
\xi^2=\frac{k}{2\beta\psi_0}\,,
%\qquad
~~~\text{and}~~~
\lambda_L^2=\frac{1}{8\pi ke^2\psi_0^2}\,.
}
The Ginzburg-Landau parameter, $\kappa=\lambda_L/\xi$, is given by 
the ratio of these characteristic length scales.

Note that since the parity-violating term in the Ginzburg-Landau model 
\Eqref{Eq:FreeEnergy} is not positively defined, the strength of the parity 
violation cannot be arbitrarily large. For the free energy to be bounded from below 
in the ground state, the parity-odd parameter $\gamma$ cannot exceed a critical value,
\Equation{Eq:Bounded:Criterion}{
0\leqslant \gamma < \gamma_\star\,,~~~\text{where}~~~ 
\gamma_\star = 
\sqrt{\frac{k}{8\pi e^2\psi_0^2}} = 
k\lambda_L\,.
}
A detailed discussion of the positive definiteness, and the derivation of the
range of validity are given in Appendix~\ref{Sec:Boundedness}.
The bound \Eqref{Eq:Bounded:Criterion} implies that the parity breaking should not 
be too strong in order to ensure the validity of the minimalistic Ginzburg-Landau 
model \Eqref{Eq:FreeEnergy}. Note, however, that the upper bound on the parity-violating 
coupling applies only to the form of the free-energy functional \Eqref{Eq:FreeEnergy}. 
If the parity-violating coupling $\gamma$ exceeds the critical value  
\Eqref{Eq:Bounded:Criterion}, the model has to be supplemented with higher-order 
terms, for the energy to be bounded.
The Lifshitz invariant in the free energy \Eqref{Eq:FreeEnergy} is given by a 
higher-order term that becomes gradually irrelevant as the system approaches a phase 
transition to the normal phase. In our work, we stay away from the criticality to 
highlight the importance of the Lifshitz term for the dynamics of the vortices.

%%%%%%%%%%%%%%%%%%%%%%%%%%%%%%%%%%%%%%%%%%%%%%%%%%%%%%%%%%%%%%%%%%%%%
%%%%%%%%%%%%%%%%%%%%%%%%%%%%%%%%%%%%%%%%%%%%%%%%%%%%%%%%%%%%%%%%%%%%%
%%%%%%%%%%%%%%%%%%%%%%%%%%%%%%%%%%%%%%%%%%%%%%%%%%%%%%%%%%%%%%%%%%%%%
\section{Vortices in noncentrosymmetric superconductors}
\label{Sec:Vortices}

Vortices are the elementary topological excitations in superconductors. Below, 
in the London limit, we derive vortex solutions for any values of the coupling 
$\gamma<\gamma_\star$. While the London limit is itself an interesting regime, 
it is important to verify that the overall physical picture advocated here is 
not merely an artifact of that particular approximation.
Consequently, we check that the 
results obtained in the London limit are consistent with the numerical solutions 
of the full nonlinear problem, by using the following procedure. 

First of all, since the Lifshitz invariant behaves as a scalar under rotations, 
the solutions should not depend on a particular orientation of the surface normal. 
Hence, there is no loss of generality to consider straight vortices along the $z$-axis. 
Such translationally invariant (straight) vortices are described, with all generality, 
by the two-dimensional field ansatz in the $xy$-plane (see remark 
\cite{
[{Field configurations that are invariant under the translations along the $z$-axis
respect the symmetries generated by the Killing vector $K_{(z)}=\partial/\partial z$. 
Moreover, all internal symmetries of the theory are gauged (the $U(1)$ gauge 
symmetry). It follows that there exist a gauge where the fields do not depend on $z$.  
this is rigorously demonstrated in: }]
[{}] Forgacs.Manton:80}): 
\Equation{Eq:Ansatz}{
\A=(A_x(x,y),A_y(x,y),A_z(x,y))~\text{and}~\psi=\psi(x,y) \,.
}
Next, in order to numerically investigate the properties of the vortex solutions, 
the physical degrees of freedom $\psi$ and $\A$ are discretized within a finite-element 
formulation \cite{Hecht:12}, and the Ginzburg-Landau free energy \Eqref{Eq:FreeEnergy} 
is subsequently minimized using a non-linear conjugate gradient algorithm. Given 
a starting configuration where the condensate has a specified phase winding 
(at large distances $\psi\propto\Exp{i\theta}$ and $\theta$ is the polar angle 
relative to the vortex center), the minimization procedure leads, after convergence 
of the algorithm, to the vortex solution of the full nonlinear theory
\cite{
[{Being in zero external field, the vortex is created only by the initial phase 
winding configuration. For further details on the numerical methods employed 
here, see for example related discussion in: }]
[{}] Garaud.Babaev.ea:16}.

\subsection{London limit solutions}
\label{Sec:London}

In the London limit, $\kappa \to \infty$, the superconducting condensate is 
approximated to have a constant density, $|\psi|=\psi_0$. Hence the current
now reads as $\j=2e\psi_0^2\left(\Grad\varphi-e\A\right)$. It leads to the second 
London equation that relates the magnetic field and $\j$
\Equation{Eq:London:B}{
  \B = \frac{1}{e}\left(\Curl\Grad\varphi-\frac{1}{2e\psi_0^2}\Curl\j\right)\,.
}
The constant density approximation, together with Eq.~\Eqref{Eq:London:B},
is then used to rewrite the Amp\`ere-Maxwell equation~\Eqref{Eq:EOM:AM} 
as the London equation for the current:
\Equation{Eq:London:J0}{
\lambda_L^2\Curl\Curl\j+\j-2\frac{\gamma}{k}\Curl\j= S\,,
}
where the source term on the right hand side reads 
\Align{Eq:London:Source}{
S&=\frac{1}{4\pi ke}\Big(\Curl\Curl\Grad\varphi 
	-\frac{\gamma}{k\lambda_L^2}\Curl\Grad\varphi\Big) \nonumber\\
&= \frac{\Phi_0}{4\pi k}\Big(\Curl\vv -
	\frac{\gamma}{k\lambda_L^2}\vv\Big) \,,
~\text{with}~\vv=\frac{1}{2\pi}\Curl\Grad\varphi\,.
}
Here $\Phi_0=2\pi/e$ is the elementary flux quantum, and $\vv$ is the density of 
vortex field that accounts for the phase singularities.

In the dimensionless units, $\tx=\frac{\x}{\lambda_L}$, $\tGrad=\lambda_L\Grad$,
the London equation is 
\Align{Eq:London:J}{
&\tCurl\tCurl\j+\j-2\Gamma\tCurl\j= 
\frac{\Phi_0}{4\pi k\lambda_L} \Big(\tCurl\vv -
	\Gamma\vv\Big) \nonumber\\%\,.
&~~\text{and}~~\B(\tx)=\Phi_0\vv-4\pi k\lambda_L\tCurl\j \,.
}
For the energy to be bounded, the criterion \Eqref{Eq:Bounded:Criterion} implies
that the dimensionless coupling $\Gamma=\gamma/k\lambda_L$ introduced here,
satisfies $0\leqslant\Gamma<1$. Defining the amplitude 
$\mathcal{A}=\frac{\Phi_0}{4\pi k\lambda_L}$, the momentum space London equation 
reads
\Align{Eq:London:Jp}{
&-\p\times\p\times\j_\p+\j_\p-2i\Gamma\p\times\j_\p= 
\mathcal{A} \Big(i\p\times\vv_\p -
	\Gamma\vv_\p\Big),
 \nonumber\\%\,,
&~~\text{and}~~\B_\p=\Phi_0\vv-4\pi i k\lambda_L\p\times\j_\p \,.
}
where $\j_\p$ is the Fourier component of the current $\j$ in the space of the 
dimensionless momenta $\p$:
\Equation{Eq:Fourier}{
\j(\tx) = \int\!\frac{d^3\p}{(2\pi)^3}\,\Exp{i\p\cdot\tx}\j_\p\,.
%\vv(\tx) = \int\!\frac{d^3\p}{(2\pi)^3}\,\Exp{i\p\cdot\tx}\vv_\p\,.
}
Similarly, the quantities $\vv_\p$ and $\B_\p$ are, respectively, the Fourier 
components of $\vv({\tx})$ and $\B({\tx})$. The solution of the algebraic 
equation \Eqref{Eq:London:Jp} in the momentum space is  
\Align{}{
j^m_\p&=\frac{\mathcal{A}}{\Sigma}\Big\{ 
		-\Gamma\big[(1-\p^2)\delta_{mn}+(\Omega+2)p^mp^n\big] \nonumber \\
	&~~~~~~+i(\Omega+2\Gamma^2)\epsilon_{mln}p^l\Big\}\,v^n_\p  
	\,:=\,\Phi_0\Lambda^{mn}_\p\,v^n_\p 
		\,,\label{Eq:London:Sol:Jp}  \\
B^m_\p&=\frac{\Phi_0}{\Sigma}\Big\{ 
		[1+(1-2\Gamma^2)\p^2]\delta_{mn}+(\Omega+2\Gamma^2)p^mp^n\nonumber \\
	&~~~~~~+i\Gamma(1-\p^2)\epsilon_{mln}p^l\Big\}\,v^n_\p    
	\,:=\,\Phi_0\Upsilon^{mn}_\p\,v^n_\p\,,\label{Eq:London:Sol:Bp}
}
with the polynomials $\Sigma\equiv\Sigma(\p^2)=(1+\p^2)^2-4\Gamma^2\p^2$ and 
$\Omega\equiv\Omega(\p^2)=1+\p^2-4\Gamma^2$. Here $\delta_{mn}$ and $\epsilon_{mln}$ 
are, respectively, the Kronecker and the Levi-Civita symbols, and the silent indices 
are summed over.

\begin{figure*}[!htb]
\hbox to \linewidth{ \hss
\includegraphics[width=0.9\linewidth]{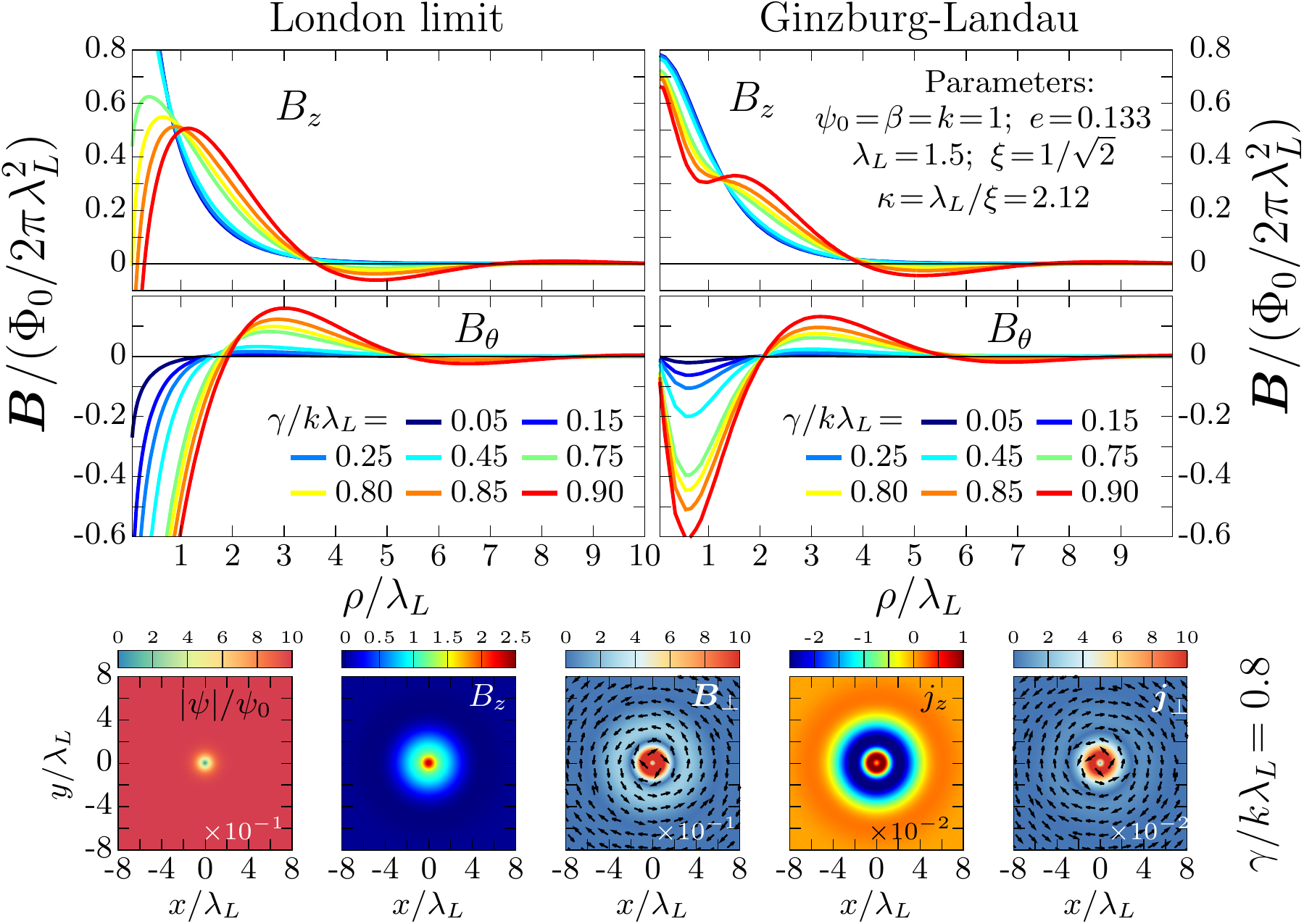}
\hss}
\caption{ 
The upper row displays the longitudinal ($B_z$) and circular ($B_\theta$) 
components of the magnetic field of a single vortex, as functions of the radial 
distance $\rho$ from the vortex center, and for various values of the parity-odd 
coupling $\gamma$. The left and right panels show the magnetic field in the 
London limit and beyond the London approximation, respectively. 
The panels in the bottom row, result from the minimization of the Ginzburg-Landau 
free energy at the parity-breaking coupling $\gamma = 0.8 \gamma_\star$. They show 
the superconducting condensate $|\psi|$, the longitudinal and transverse components 
of the magnetic field, $B_z$ and ${\bs B}_\perp$, and currents , $j_z$ and 
$\j_\perp$, in the transverse plane of the vortex.
In the case of a weak parity violation, $\gamma\ll\gamma_\star$, the longitudinal 
component of the magnetic field is similar to that of conventional Abrikosov vortices 
for which $B_z(\rho)$ is monotonic and exponentially localized around the vortex core at 
$\rho = 0$.
When the parity-breaking term becomes large, with $\gamma$ approaching the critical 
value $\gamma_\star$, the longitudinal component $B_z$ becomes a non-monotonic 
function as the distance $\rho$ from the vortex core increases.
}
\label{Fig:MagneticField}
\end{figure*}

Thus, the vortex field $\vv$ completely determines, via its Fourier image $\vv_\p$, 
the momentum-space representations of the current \Eqref{Eq:London:Sol:Jp} and of 
the magnetic field \Eqref{Eq:London:Sol:Bp}. The corresponding real-space solutions 
are obtained by the Fourier transformation~\Eqref{Eq:Fourier}.
Assuming the translation invariance along the $z$-axis, a set of $N$ vortices located 
at the positions $\tx_a$, and characterized by the individual winding numbers $n_a$ 
(with $a=1,\cdots,N$), is described by the Fourier components 
\Equation{Eq:London:Ansatz}{
\vv_\p= 2\pi\frac{\delta (p_z){\bs{\mathrm{e}}}_z}{\lambda_L^2}
\sum_{a=1}^Nn_a\Exp{-i\p\cdot\tx_a}\,,
}
where the Dirac delta for the momentum $p_z$ specifies the translation invariance 
of the configuration.

\subsection{Single vortex}

The analysis becomes particularly simple for a single elementary vortex with a 
unit winding number ($n_1{=}1$) located at the origin ($\x_1 = 0$). The 
corresponding magnetic field reads as follows:
\Equation{Eq:London:Vortex:B}{
\B_\p=\frac{2\pi\Phi_0\delta (p_z)}{\lambda_L^2\Sigma}
\Vector{\phantom{-}i\Gamma(1-\p^2)p_y     \\
		         - i\Gamma(1-\p^2)p_x     \\
		(1-2\Gamma^2)\p^2+1
		}.
}
Next, we express the position, $\tx=(\trho\cos\theta,\trho\sin\theta,\tz)$, and 
momentum, $\p=(q\cos\vartheta,q\sin\vartheta,p_z)$, in cylindrical coordinates.
An integration over the angular degrees of freedom $\vartheta$ nullifies the 
radial part $B_\rho$ of the magnetic field and generates the Bessel functions of 
the first kind, $J_0$ and $J_1$. Hence, the nonzero components of the magnetic 
field can be expressed as one-dimensional integrals over the radial momentum $q$:
\Align{Eq:London:B:sol:0}{
&B_\theta\Big(\frac{\rho}{\lambda_L}\Big)\!=\! \frac{\Phi_0\Gamma}{2\pi\lambda_L^2}
	\int_0^\infty\! 
		\frac{q^2(1-q^2)dq}{(1+q^2)^2-4\Gamma^2q^2}
		J_1\Big(\frac{q\rho}{\lambda_L}\Big) %\label{Eq:London:Btheta}
		\,,\nonumber\\
&B_z\Big(\frac{\rho}{\lambda_L}\Big)\!\!=\! \frac{\Phi_0}{2\pi\lambda_L^2}\!
	\int_0^\infty\! \!
		\frac{q[(1-2\Gamma^2)q^2+1]dq}{(1+q^2)^2-4\Gamma^2q^2}
		J_0\Big(\frac{q\rho}{\lambda_L}\Big).%\label{Eq:London:Bz}
}
Similarly, the nonzero components of the current are:
\Align{Eq:London:J:sol:0}{
&j_\theta\Big(\frac{\rho}{\lambda_L}\Big)\!=\! \frac{\Phi_0}{8\pi^2k\lambda_L^3}
	\int_0^\infty\! 
		\frac{q^2(q^2+1-2\Gamma^2)dq}{(1+q^2)^2-4\Gamma^2q^2}
		J_1\Big(\frac{q\rho}{\lambda_L}\Big) %\label{Eq:London:Jtheta}
		\,,\nonumber\\
&j_z\Big(\frac{\rho}{\lambda_L}\Big)\!\!=\! \frac{-\Phi_0\Gamma}{8\pi^2k\lambda_L^3}\!
	\int_0^\infty\! \!
		\frac{q(1-q^2)dq}{(1+q^2)^2-4\Gamma^2q^2}
		J_0\Big(\frac{q\rho}{\lambda_L}\Big).%\label{Eq:London:Jz}
}
Using the Hankel transform \cite{Piessens:18}, as demonstrated in detail in Appendix 
\ref{Sec:Integrals}, these integrals can be solved analytically in terms of the 
modified Bessel functions of the second kind $K_\nu$. Introducing the complex number 
$\eta=\Gamma-i\sqrt{1-\Gamma^2}$, the nonzero components of the magnetic field read 
\Align{Eq:London:B:sol}{
&B_\theta\Big(\frac{\rho}{\lambda_L}\Big)\!=\! \frac{\Phi_0}{2\pi\lambda_L^2}
	\Re\left[i\eta^2 K_1\left(\frac{i\eta\rho}{\lambda_L}\right) \right] \,,\nonumber\\
&B_z\Big(\frac{\rho}{\lambda_L}\Big)\!\!=\! \frac{-\Phi_0}{2\pi\lambda_L^2}\!
	\Re\left[\eta^2 K_0\left(\frac{i\eta\rho}{\lambda_L}\right) \right]\,.
}
Similarly, the nonzero components of $\j$ are:
\Align{Eq:London:J:sol}{
&j_\theta\Big(\frac{\rho}{\lambda_L}\Big)\!=\! \frac{-\Phi_0}{8\pi^2k\lambda_L^3}
	\Re\left[i\eta K_1\left(\frac{i\eta\rho}{\lambda_L}\right) \right]	\,,\nonumber\\
&j_z\Big(\frac{\rho}{\lambda_L}\Big)\!\!=\! \frac{\Phi_0}{8\pi^2k\lambda_L^3}\!
	\Re\left[\eta K_0\left(\frac{i\eta\rho}{\lambda_L}\right) \right]\,.
}
In the absence of parity breaking ($\Gamma=0$ and $\eta=-i$), the above integrals 
expectedly give the textbook expressions for the nonvanishing components of the 
magnetic field, 
$B_z(\rho/\lambda_L)\!=\!\frac{\Phi_0}{2\pi\lambda_L^2} K_0(\rho/\lambda_L)$, 
and of the superconducting current,
$4\pi kj_\theta(\rho/\lambda_L)\!=\!\frac{\Phi_0}{2\pi\lambda_L^3}K_1(\rho/\lambda_L)$.

Figure \ref{Fig:MagneticField} shows the magnetic field of a single vortex both 
in the London limit and for the full Ginzburg-Landau problem. First, although the 
solutions are expected to differ at the vortex core, the overall behavior remains 
qualitatively similar in both cases. Indeed, the London solutions are divergent at 
the vortex core, and thus they require a sharp cut-off at the coherence length~$\xi$. 
Solutions beyond the London limit, on the other hand, are regular everywhere.  
The bottom row of \Figref{Fig:MagneticField} shows a typical vortex solution 
obtained numerically beyond the London limit. This is a close-up view of the 
vortex core structure, while the actual numerical domain is much larger in 
order to prevent any finite-size effect. 
While the density profile is similar to that of common vortices, the magnetic 
field shows a pretty unusual profile featuring a slight inversion, away from 
the center. For the current parameter set, where $\gamma=0.8\gamma_\star$, the 
amplitude of the reversed field compared to the maximal amplitude is rather 
small. Yet, as illustrated on the top-right panels of \Figref{Fig:MagneticField}, 
the amplitude of inversion of the magnetic field, typically increases  with the 
parity-breaking coupling $\gamma$. Thus when $\gamma$ is close to the critical 
coupling $\gamma_\star$, the magnitude of the responses and field inversions
become more important.

When the parity-breaking coupling $\gamma$ is small compared to the upper bound  
$\gamma_\star$, the longitudinal component $B_z$ of the magnetic field is 
monotonic and exponentially localized, as for conventional vortices. The vortex 
configurations start to deviate from the conventional case when the parity 
breaking strengthens. Indeed, when $\gamma$ increases, the magnetic field $B_z$ 
does not vary monotonically any longer. As can be seen in the top-right panel of 
\Figref{Fig:MagneticField}, it can be reversed, and even features several local minima 
as $\gamma$ approaches its critical value $\gamma_\star$. Note that, the complicated 
spatial structure and inversion of the magnetic field also comes with the inversion 
of the supercurrents.
The distance from the vortex center $\rho \simeq 4\lambda_L$, where the longitudinal 
component of the magnetic field first vanishes, corresponds to the radius where the 
in-plane current $j_\theta$ reverses its sign. Similarly, the longitudinal current 
$j_z$ vanishes for the first time at the shorter distance to the vortex core, 
$\rho \simeq 2\lambda_L$, where the circular magnetic field cancels, $B_\theta=0$. 
These observations are consistent with the results from the perturbative regime, 
$\gamma\ll\gamma_\star$~\cite{Kashyap.Agterberg:13}.
Interestingly, these specific radii are pretty much unaffected by the 
value of the parity-breaking coupling. 

Note that the structure of the zeros of the modified Bessel functions with 
complex arguments shows that any non-zero value of the parity-breaking coupling 
$\gamma$ exhibits zeros at some distance away from the singularity. In practice, 
for small values of $\gamma$, the first zero is pushed very far from the vortex 
core, and the amplitude of the field inversion is vanishingly small. Hence while 
the field inversion formally occurs at all finite $\gamma$, it becomes noticeable 
when $\gamma$ is not too small. The structure of magnetic field inversion as a function 
of $\gamma$ qualitatively resembles the alternating attractive/positive regions 
displayed in the right panel of \Figref{Fig:Interactions}.

\section{Vortex interactions}
\label{Sec:Interactions}

\begin{figure*}[!htb]
\hbox to \linewidth{ \hss
\includegraphics[width=0.9\linewidth]{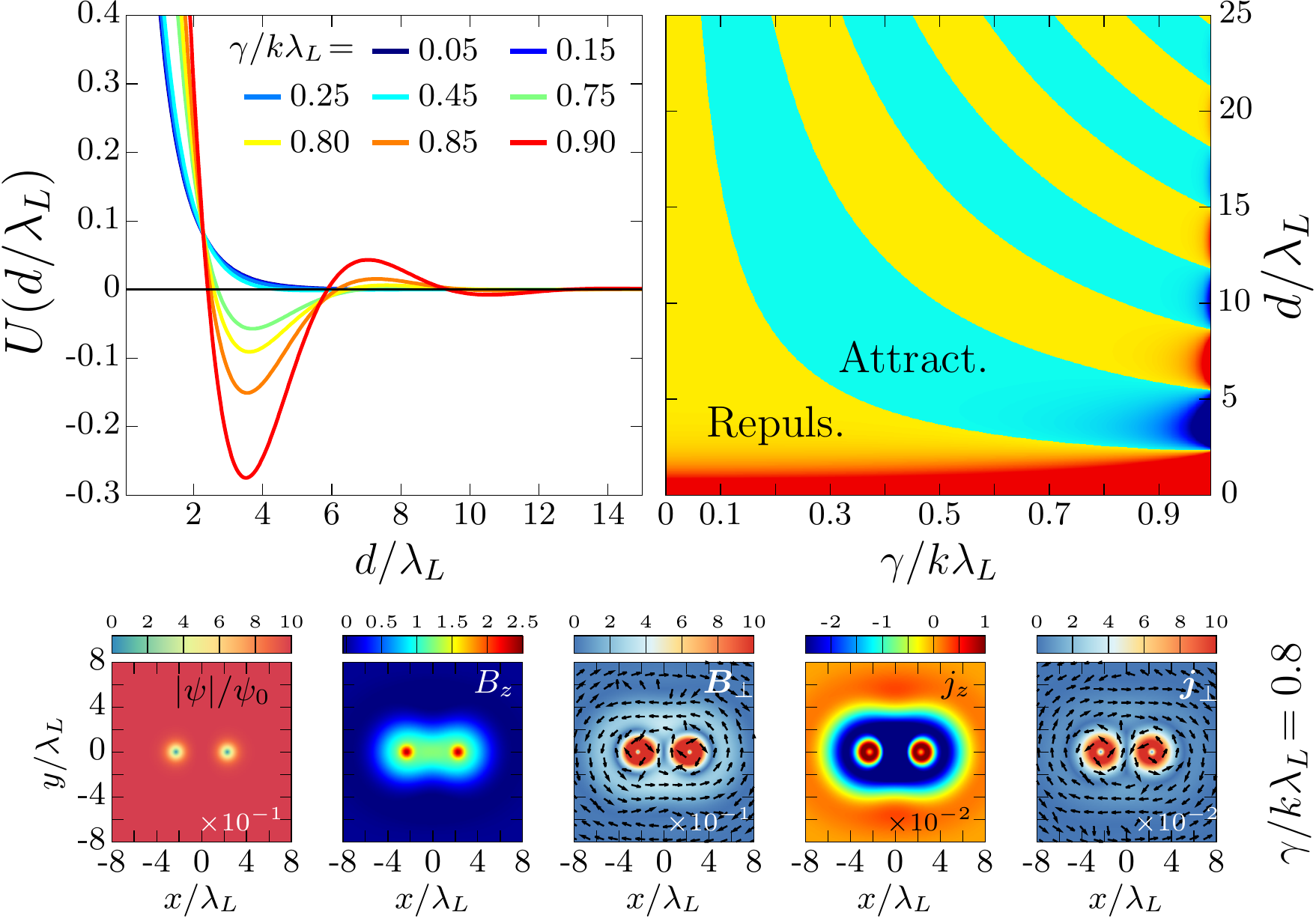}
\hss}
\caption{ 
The top-left panel shows the function $U(d/\lambda_L)$ that controls the intervortex 
interactions, as a function of the distance $d$ between the vortices, for various 
values of the parity-odd coupling $\gamma$. 
The top-right panel displays a phase diagram showing the attractive (blue) and 
repulsive (yellow) regions depending on the parity breaking coupling $\gamma$ and 
the intervortex distance $d$. The panels in the top row corresponds to the London limit.
The panels on the bottom row, obtained within the Ginzburg-Landau calculation, show 
various physical quantities in the transverse plane of a vortex bound state, for the 
parity-breaking coupling $\gamma = 0.8 \gamma_\star$ (the other parameters are the 
same as in \Figref{Fig:MagneticField}). This vortex pair, obtained after convergence 
of minimization of the Ginzburg-Landau free energy, demonstrates that the property 
of the non-monotonic interactions can survive beyond the London limit. The London limit 
estimates the intervortex separation $d_{LL}=3.6\lambda_L$, while the Ginzburg-Landau 
calculation finds $d_{GL}=4.6\lambda_L$. Despite the fact that the GL simulation is far 
from the London limit (here $\kappa=2.12$), both values are in qualitative agreement.
}
\label{Fig:Interactions}
\end{figure*}

The possibility of having an inversion of the magnetic field suggests that the 
interaction between two vortices might be much more involved than the pure 
repulsion that occurs in conventional type-2 superconductors. Indeed, since the 
conventional long-range intervortex repulsion is due to the magnetic field, 
it is quite likely that the interaction here might be not only quantitatively, 
but also qualitatively altered.
To investigate these, we consider the London limit free energy $F$ written in the 
previously used dimensionless coordinates. Using Eq.~\Eqref{Eq:Fourier} 
to express the quantities $\j$ and $\B$ in terms of their Fourier components,
we find the expression of the free energy in the momentum space:
\Align{Eq:FreeEnergy:K1}{
F=\frac{\lambda_L^3}{8\pi}\int \frac{d^3\p}{(2\pi)^3}\Big\{ &\B_\p\cdot\B_{-\p}
+(4\pi k\lambda_L\j_\p)\cdot(4\pi k\lambda_L\j_{-\p}) \nonumber \\
&+2\Gamma(4\pi k\lambda_L\j_\p)\cdot\B_{-\p}\Big\}\,.
}
Replacing the Fourier components of the magnetic field $\B_\p$ and of the current 
$\j_\p$ with the corresponding expressions in terms of the vortex field $\vv_\p$, 
Eqs.~\Eqref{Eq:London:Sol:Jp} and \Eqref{Eq:London:Sol:Bp}, respectively,
yields the free energy: 
\Align{Eq:FreeEnergy:K2}{
&~~~~F=\frac{\lambda_L^3}{8\pi}\int \frac{d^3\p}{(2\pi)^3}\,
	\vv_\p^m G^{mn}\vv_{-\p}^n \\
&\text{where}~G^{mn}=\Upsilon^{lm}_\p\Upsilon^{ln}_{-\p} +
\Lambda^{lm}_\p\Lambda^{ln}_{-\p} 
+2\Gamma\Lambda^{lm}_\p\Upsilon^{ln}_{-\p} \,. \nonumber
} 
The interaction matrix $G$ has a rather involved structure. Yet, given that 
only the axial Fourier components of the vortex field \Eqref{Eq:London:Ansatz} 
are nonzero, only the component $G^{zz}$ will contribute to the energy.
Up to terms that are proportional to $p_z$, and thus will be suppressed by 
the Dirac delta $\delta(p_z)$, $G^{zz}$ takes the simple form 
\Equation{Eq:Gzz}{
G^{zz}=\frac{(1-\Gamma^2)(1+\p^2)}{\Sigma}~+~(\text{terms}\propto p_z) \,.
}
Finally, using the vortex field ansatz \Eqref{Eq:London:Ansatz}, together with 
the expressions \Eqref{Eq:FreeEnergy:K2} and \Eqref{Eq:Gzz} determines the free 
energy associated with a set of translationally invariant vortices 
\Equation{Eq:FreeEnergy:K3}{
F=\frac{\Phi_0^2(1{-}\Gamma^2)}{8\pi\lambda_L}\sum_{a,b=1}^N n_an_b\!\int\! 
\frac{d^2\p}{2\pi}\frac{1+\p^2}{\Sigma(\p^2)}\Exp{i\p\cdot(\tx_a-\tx_b)}.
}
The two dimensional integration in \Eqref{Eq:FreeEnergy:K3} can further be 
simplified and finally, the free energy reads as:
\Align{Eq:FreeEnergy:K4}{
F=\frac{\Phi_0^2(1-\Gamma^2)}{8\pi\lambda_L}&\sum_{a,b=1}^Nn_an_b 
U\left(\frac{|x_a-x_b|}{\lambda_L}\right)  \\
\text{where}~~U(x)&=\int_0^\infty\! \frac{q(1+q^2)\,dq}{(1+q^2)^2-4\Gamma^2q^2} 
	\,J_0(q x)\nonumber \\
&=\Re\left[\frac{i\eta}{\sqrt{1-\Gamma^2}} K_0\left(i\eta x\right) \right]
\,. \label{Eq:Interaction}
}
Hence the free energy of a set of vortices reads as:
\Equation{Eq:FreeEnergy:K5}{
F=F_0+\frac{\Phi_0^2(1-\Gamma^2)}{4\pi\lambda_L}\sum_{
a,b>a}^N n_an_b
U\left(\frac{|x_a-x_b|}{\lambda_L}\right)\,,
}
where the term $F_0=\frac{\Phi_0^2(1-\Gamma^2)}{8\pi\lambda_L}\sum_an_a^2U(\xi)$ 
accounts for the self-energy of individual vortices. Since $U(x)$ diverges at small 
separations $x$, the self-energy has to be regularized at the coherence length 
$\xi\ll\lambda_L$, which determines the size of the vortex core. 
The interaction energy of the vortices separated by a distance $d$ is thus determined 
by the function $U(d/\lambda_L)$. In the absence of parity breaking ($\Gamma=0$ and 
$\eta=-i$), Eq.~\eq{Eq:FreeEnergy:K5} leads again to the textbook expression for the 
interaction energy $V_\mathrm{int}(d/\lambda_L)=\Phi_0^2K_0(d/\lambda_L)/4\pi\lambda_L$.

Figure \ref{Fig:Interactions} displays the function $U(d/\lambda_L)$, which controls 
the interacting potential between vortices, calculated in the London limit. 
For vanishing $\gamma$, the interaction is purely repulsive, and it is altered by 
a nonzero coupling. As shown in the left panel, when increasing $\gamma/k\lambda_L$ 
the interaction can become non-monotonic with a minimum at a finite distance of 
about $4\lambda_L$. Upon further increase of the coupling $\gamma$ toward the 
critical coupling $\gamma_\star$, the interacting potential can even develop 
several local minima. The phase diagram on the right panel of 
\Figref{Fig:Interactions} shows the different attractive and repulsive 
regions as functions of~$\gamma$ and of the vortex separation.

The fact that the interaction energy features a minimum at a finite distance implies 
that a pair of vortices tends to form a bound state. As can be seen in the bottom 
row of \Figref{Fig:Interactions}, the tendency of vortices to form bound-states
persists beyond the London approximation. This configuration is obtained numerically 
by minimizing the Ginzburg-Landau free energy \Eqref{Eq:FreeEnergy}. Notice that these 
bottom panels show a close-up view of the vortex pair, while the actual numerical domain 
is much larger 
\footnote{
We also performed numerical simulations for three and four vortices and observed 
that they form bound vortex clusters as well.
}. 
The formation of a vortex bound state can heuristically be understood as a compromise 
between the axial magnetic repulsion of $B_z$ which competes with in-plane attraction 
mediated by $B_\perp$. The bound-state formation can alternatively be understood to 
originate from the competition between the in-plane and axial contributions of the 
currents. First of all, the in-plane screening currents mediate, as usual, repulsion 
between vortices. The interaction between the axial components of the currents, on the 
other hand, mediates an attraction, just like the force between parallel wires carrying 
co-directed electric currents. 

The non-monotonic behavior of the magnetic field and currents thus leads to 
non-monotonic intervortex interactions, and therefore allows for bound-states of 
vortices or clusters to form. Such a situation is known to exist in multicomponent 
superconductors due to the competition between various length scales 
(see e.g. \cite{Babaev.Speight:05,Babaev.Carlstrom.ea:12,Carlstrom.Garaud.ea:11,
Babaev.Carlstrom.ea:17,Silaev.Winyard.ea:18} as well as \cite{Haber.Schmitt:17,
Haber.Schmitt:18} for superconducting/superfluid systems). In an applied external 
field, the existence of non-monotonic interactions allows for a macroscopic phase 
separation into domains of vortex clusters and vortex-less Meissner domains. 
The situation here contrasts with the multicomponent case, as it occurs only due 
to the existence of Lifshitz invariants. 
In two-dimensional systems of interacting particles, multi-scale potentials and 
non-monotonic interactions are known to be responsible for the formation of rich 
hierarchical structures. These structures include clusters of clusters, concentric 
rings, clusters inside a ring, or stripes \cite{OlsonReichhardt.Reichhardt.ea:10,
Varney.Sellin.ea:13}. It can thus be expected that very rich structures would appear 
in noncentrosymmetric superconductors as well. However, a verification of this 
conjecture is beyond the scope of the current work, as it deserves a separate 
detailed investigation.

As shown in \Figref{Fig:Interactions}, the interaction energy 
$V_{v/v}(x)\propto U(x)$ between two vortices with unit winding $n_1=n_2=1$ can 
thus lead to the formation of a vortex bound state. A very interesting property 
is that it also opens the possibility of vortex/anti-vortex bound-states. 
Indeed, according to Eq.~\Eqref{Eq:FreeEnergy:K5} the interaction of a vortex 
$n_1=1$ and an anti-vortex $n_2=-1$ corresponds to a reversal of the interacting 
potential: $V_{v/av}(x)\propto-U(x)$. Thus from \Figref{Fig:Interactions} it is 
clear that if a vortex/anti-vortex pair is small enough, it will collapse to zero 
and thus lead to the vortex/anti-vortex annihilation.
Now, considering for example the curve $\gamma/k\lambda_L=0.9$ in 
\Figref{Fig:Interactions}, if the size of the vortex/anti-vortex pair is larger 
than $4\lambda_L$, there exists an energy barrier that prevents the pair from further 
collapse. Hence the vortex/anti-vortex pair should relax to a local minimum of the 
interaction energy. The resulting vortex/anti-vortex bound state has thus a size of 
approximately $7\lambda_L$. 
Note that the above analysis demonstrates that vortex/anti-vortex bound-states do 
exist as meta-stable states in the London limit. It is quite likely that these results 
are still qualitatively valid beyond the London approximation at least for strong 
type-2 superconductors. The possibility to realize vortex/anti-vortex pairs for 
weakly type-2 superconductors requires careful analysis and is beyond the scope of the 
current work.

%%%%%%%%%%%%%%%%%%%%%%%%%%%%%%%%%%%%%%%%%%%%%%%%%%%%%%%%%%%%%%%%%%%%%
\section{Conclusions}\label{Conclusion}

In this paper we have demonstrated that the vortices in noncentrosymmetric cubic 
superconductors feature unusual properties induced by the possible reversal of the 
magnetic field around them. Indeed, the longitudinal (i.e., parallel to the vortex 
line) component of the magnetic field changes sign at a certain distance away from 
the vortex core. Contrary to the vortices in a conventional superconductor, the 
magnetic-field reversal in the parity-broken superconductor leads to non-monotonic 
intervortex forces which can act both attractively and repulsively depending on 
the distance separating individual vortices.

These properties have been demonstrated analytically within the London limit. 
Our numerical analysis of the nonlinear Ginzburg-Landau theory proves that the 
magnetic-field reversal and the non-monotonic intervortex forces survive beyond 
the London approximation in noncentrosymmetric superconductors.

Due to the nonmonotonic intervortex interactions, the vortices in the 
parity-breaking superconductors may form unusual states of vortex matter,  
such as bound states and clusters of vortices. The structure of the interaction  
potential strongly suggests that very rich vortex matter structures can emerge.  
For example, hierarchically structured quasi-regular vortex clusters, stripes 
and more, are typical features of the interacting multi-scale and non-monotonic 
interaction potentials \cite{OlsonReichhardt.Reichhardt.ea:10,Varney.Sellin.ea:13}. 
Moreover, given the possibility to form vortex/anti-vortex bound states, we can 
anticipate important consequences for the statistical properties and phase 
transitions in such models.

\emph{Note added:} 
In the process of completion of this work, we were informed about an independent 
work by Samoilenka and Babaev \cite{Samoilenka.Babaev:20} showing similar results 
about vortices and their interactions. The submission of this work was coordinated 
with that of \cite{Samoilenka.Babaev:20}.

%%%%%%%%%%%%%%%%%%%%%%%%%%%%%%%%%%%%%%%%%%%%%%%%%%%%%%%%%%%%%%%%%%%%%
\begin{acknowledgments}
We acknowledge fruitful discussions with D.~F.~Agterberg, E.~Babaev and F.~N.~Rybakov 
and A.~Samoilenka. We especially thank A.~Samoilenka for suggesting to use the 
Hankel transform in our analytical calculations. 
The work of M.C. was partially supported by Grant No. 0657-2020-0015 of the Ministry 
of Science and Higher Education of Russia.
The work of D.K. was supported by the U.S. Department of Energy, Office of Nuclear 
Physics, under contracts DE-FG-88ER40388 and DE-AC02-98CH10886, and by the Office 
of Basic Energy Science under contract DE-SC-0017662.
The computations were performed on resources provided by the Swedish 
National Infrastructure for Computing (SNIC) at National Supercomputer 
Center at Link\"{o}ping, Sweden. 

\end{acknowledgments}

%%%%%%%%%%%%%%%%%%%%%%%%%%%%%%%%%%%%%%%%%%%%%%%%%%%%%%%%%%%%%%%%%%%%%
%%%%%%%%%%%%%%%%%%%%%%%%%%%%%%%%%%%%%%%%%%%%%%%%%%%%%%%%%%%%%%%%%%%%%
%%%% Bibliography
%%%\bibliographystyle{apsrev4-1}
%\bibliography{../Non-centrosymmetric}

\clearpage
\appendix
\setcounter{section}{0}
\setcounter{equation}{0}
\renewcommand{\theequation}{\Alph{section}\arabic{equation}}

\section{Positive definiteness of the energy}
\label{Sec:Boundedness}

The free energy \Eqref{Eq:FreeEnergy} should be bounded from below in order to 
be able to describe the ground state of the NCS superconductor. To demonstrate 
the boundedness, we use the relations 
\SubAlign{EqApp:Bounded:Relations}{
\j&=2e|\psi|^2\left(\Grad\varphi-e\A\right),  \\
|\D\psi|^2&=\left(\Grad|\psi|\right)^2+|\psi|^2\left(\Grad\varphi-e\A\right)^2\, \\
	&=\left(\Grad|\psi|\right)^2+\frac{|\j|^2}{4e^2|\psi|^2}\,,
}
to rewrite the energy density in the following form:
\SubAlign{EqApp:Bounded:FreeEnergy}{
 \F&= \frac{\B^2}{8\pi}+k\left(\Grad|\psi|\right)^2+\frac{k|\j|^2}{4e^2|\psi|^2}
 +\gamma\j\cdot\B+V[\psi]   	\\% +\frac{\beta}{2}(|\psi|^2-\psi_0^2)^2 
 &=\frac{1}{8\pi}\left[\B^2 + 8\pi\gamma\j\cdot\B\right] +\frac{k|\j|^2}{4e^2|\psi|^2}
 	\nonumber\\
 	&~~~~~~+k\left(\Grad|\psi|\right)^2+\frac{\beta}{2}(|\psi|^2-\psi_0^2)^2	\\
 &=\frac{1}{8\pi}\big|\B+4\pi\gamma\j\big|^2 
 	+\Big(\frac{k}{4e^2|\psi|^2}-2\pi\gamma^2\Big)|\j|^2\nonumber\\
 	&~~~~~~+k\left(\Grad|\psi|\right)^2+\frac{\beta}{2}(|\psi|^2-\psi_0^2)^2	\,.
\label{EqApp:Bounded:FreeEnergy:3}
}
Leaving aside all terms with the perfect squares in 
Eq.~\Eqref{EqApp:Bounded:FreeEnergy:3}, we find that the only criterion for the free 
energy to be bounded from below is to require the prefactor in front of the $|\j|^2$ 
term to be positive. We arrive at the following condition of the stability of the 
system~\Eqref{Eq:FreeEnergy}:
\Equation{EqApp:Bounded:Criterion:1}{
\gamma^2<\frac{k}{8\pi e^2|\psi|^2}\,.
}

In the ground state with $|\psi|=\psi_0$, the stability condition
\Eqref{EqApp:Bounded:Criterion:1} reduces to the simple inequality:
\Equation{EqApp:Bounded:Criterion:2}{
\gamma<\gamma_\star=k\lambda_L\,.
}
where $\lambda_L$ is the London penetration depth~\eq{Eq:LengthScales}. 

In the London limit, the superconducting density $|\psi|^2$ is a fixed constant 
quantity regardless of the external conditions. Therefore, the Ginzburg-Landau theory 
for the NCS superconductor in the London limit is always bounded from below provided 
the Lifshitz--invariant coupling $\gamma$ satisfies Eq.~\Eqref{EqApp:Bounded:Criterion:2}.

\subsection*{Positive definiteness beyond the London limit}

The issue of positive definiteness is less obvious beyond the London limit. 
Indeed, let us first assume that the values of the parameters $(e, k, \gamma)$ are 
chosen in such a way that the formal criterium~\eq{EqApp:Bounded:Criterion:1} is satisfied. 
If we neglect the fluctuations of the condensate $\psi$ (this requirement is always 
satisfied in the London regime) then we indeed find that the ground state resides in 
a locally stable regime so that all terms in the free energy 
\Eqref{EqApp:Bounded:FreeEnergy:3} are positively defined. However, the density $|\psi|$ 
is, in principle, allowed to take any value, and large enough fluctuations of $|\psi|$ 
might trigger an instability. A possible signature of the instability can indeed be 
spotted in the property that a variation of the absolute value of the condensate about 
the ground state, $|\psi| = \psi_0 + \delta |\psi|$, gives a negative contribution to 
the free energy, $\delta F = - k \delta |\psi| /(2 e^2 \psi_0^3)$, in the linear order, 
provided all other parameters are fixed. 

To illustrate a possible mechanism of the development of the instability 
inside the noncentrosymmetric superconductor, let us consider a large enough local 
region characterized by a uniform, coordinate-independent condensate $\psi$. For this 
configuration, the third (gradient) term in the free-energy density
\Eqref{EqApp:Bounded:FreeEnergy:3} is identically zero. Gradually increasing the value 
of the condensate beyond the ground state value $\psi_0$, we increase the fourth 
(potential) term in Eq.~\eq{EqApp:Bounded:FreeEnergy:3}, which make this change 
energetically unfavorable. On the other hand, as the condensate crosses the threshold 
of applicability of Eq.~\eq{EqApp:Bounded:Criterion:1}, then the second term in 
the free energy~\eq{EqApp:Bounded:FreeEnergy:3} becomes negatively defined, and the 
development of the current ${\j}$ leads to the unbounded decrease of this term. The rise 
in the current $\j$ will, in turn, affect the first (magnetic) term, which may be 
compensated by a rearranging of the magnetic field $\B$ with the local environment in 
such a way that the combination $\B+4\pi\gamma\j$ keeps a small value in the discussed 
region.

\begin{figure*}[!htb]
\hbox to \linewidth{ \hss
\includegraphics[width=0.9\linewidth]{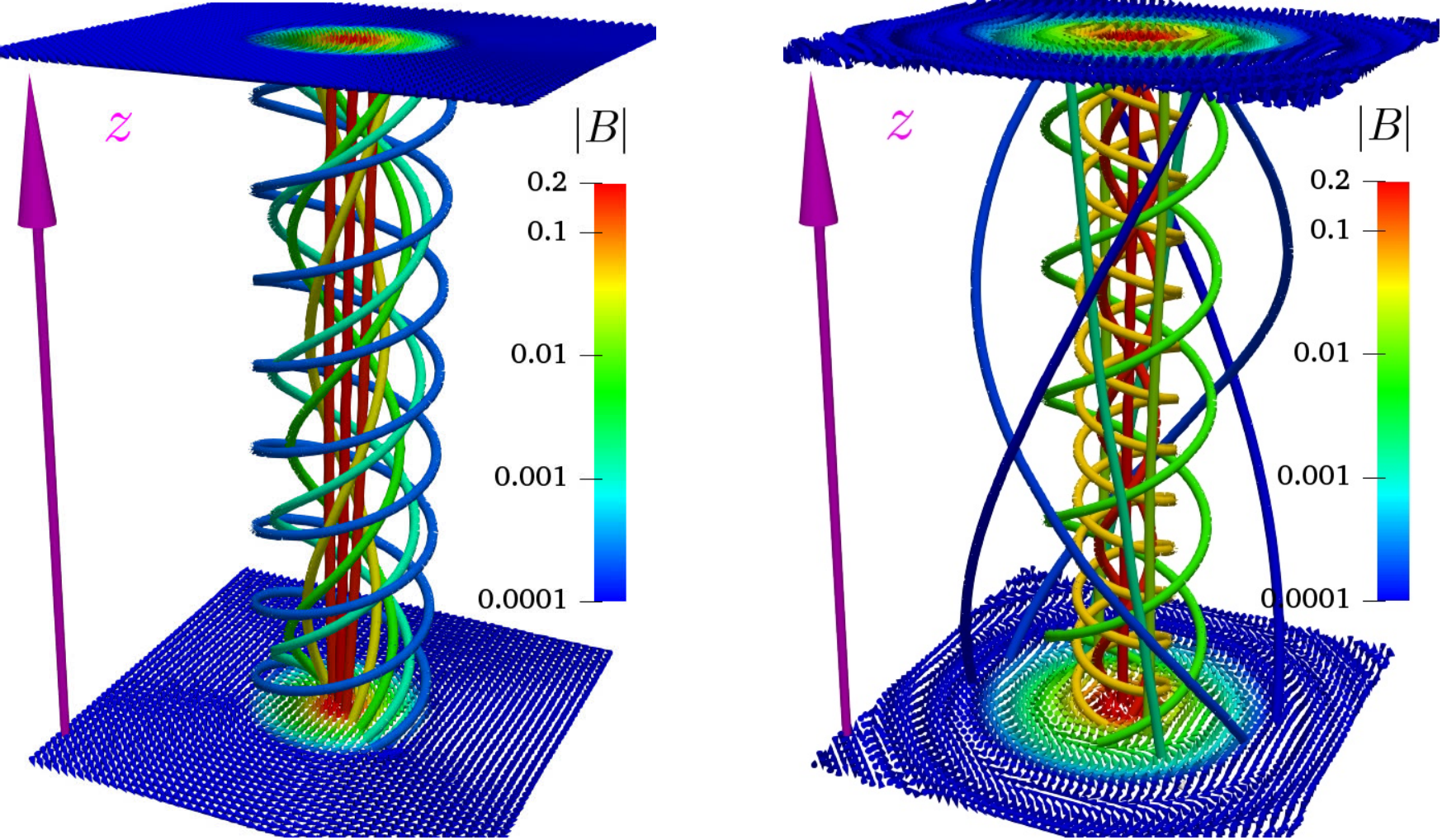}
\hss}
\caption{ 
Helical structure of the magnetic field streamlines around vortices in 
noncentrosymmetric superconductors. The magnetic field is displayed on the two 
planes normal with respect to the vortex line. The colors encode the amplitude 
$|B|$, while the arrows demonstrate the orientation of the field. The tubes 
represent streamlines of the magnetic field between both planes. 
The left panel shows the helical structure of the streamlines for a moderate value 
of the parity-breaking coupling, $\gamma/k\lambda=0.2$. The streamlines here feature 
all the same chirality.
The right panel corresponds to rather important parity-breaking coupling 
$\gamma/k\lambda=0.8$, for which the longitudinal component of the magnetic field 
is inverted at some distance from the core. The chirality of the streamline 
depends on whether the longitudinal component of the magnetic field is inverted.
}
\label{Fig:Helicity}
\end{figure*}

Notice that the presence of isolated vortices makes the system stable as in a vortex 
core the condensate vanishes, $\psi\to0$, and the second, potentially unbounded term 
in \Eqref{EqApp:Bounded:FreeEnergy:3} becomes positively defined. In our numerical 
simulations, we were also spotting certain unstable patterns especially in the regimes 
when the Lifshitz-invariant coupling $\gamma$ was chosen to close to the critical 
value $\gamma_\star$ in the ground state \Eqref{EqApp:Bounded:Criterion:2}. For example, 
a system of randomly placed multiple elementary vortices relaxes their free energy via 
mutual attraction and the formation of a common bound state. Since the vortex bound state 
hosts a stronger circular electric current, it becomes possible to overcome the 
stability by `compressing' the vortex cluster, and then destabilizing the whole system.

We conclude that the processes that permit fluctuations of the condensate $|\psi|$ 
towards the large values (as compared to the ground-state value $\psi_0$) could activate 
the destabilization of the whole model. Theoretically, the unboundedness of the free 
energy from below may appear to be an unwanted feature of the model. However, one 
should always keep in mind that the Ginzburg-Landau functional is a leading part of the 
gradient expansion of an effective model, and there always exist higher power gradients 
that will play a stabilizing role preventing the unboundedness from actually being realized 
in a physically relevant model. 

\section{Vortex helicity}
\label{Sec:Helicity}

As emphasized in the main body of the paper, the magnetic field of vortex states 
in cubic noncentrosymmetric superconductors features a helicoidal structure around 
the core. This is illustrated in \Figref{Fig:Helicity}, which displays a typical 
magnetic field structure around vortex cores. The magnetic field there is determined 
within the London approximation \Eqref{Eq:London:B:sol}. Each helical tube represents 
a streamline of the magnetic field which is tangent to the magnetic field in every point.
\Figref{Fig:Helicity} clearly shows that every streamline forms an helix along the 
axis $z$. Each helix has a period that depends on the distance from the helix to 
the center of the vortex. The latter property indicates that, contrary to the magnetic 
field itself, the streamlines of the magnetic field are not invariant for the 
translations along the $z$ axis.

\Figref{Fig:Helicity} shows two qualitatively different situation of moderate 
(left panel) and important (right panel) parity-breaking coupling $\gamma$.
For moderate parity-breaking coupling, the magnetic field streamlines have a
helical structure with a pitch that varies with the distance from the vortex 
core. Note that all streamlines have the same chirality, which is specified by 
the sign of the parity-breaking coupling $\gamma$.
On the other hand, as discussed in the main body of the paper, vortices feature 
inversion of the magnetic field $\B$ for important parity-breaking coupling $\gamma$. 
As a result, the chirality of the streamline depends on whether the longitudinal 
component of the magnetic field is inverted.
More details about the helical structure of the magnetic field can be seen
from animations in the supplemental material \cite{Supplemental-arxiv}.

\section{Description of the supplementary animations (see ancillary files)}
\label{Sec:Movies}

There are three animations that illustrate the results of presented in the manuscript. 
The magnetic field forms helical patterns around a straight static vortex, in a 
noncentrosymmetric superconductor. For important values of the parity-breaking coupling 
$\gamma$, the magnetic field $\B$ can further show inversion patterns around the vortex. 
That is, as the distance from the vortex core increases, the longitudinal component of the 
magnetic field may change it sign.

The supplementary animations display the following: 
On the two static planes, normal with respect to the vortex line, the colors encode 
the amplitude of the magnetic field B, while the arrows demonstrate the orientation of 
the field. The tubes represent streamlines of the magnetic magnetic field between both 
planes.

\begin{itemize}

\item \textbf{\tt movie-1.avi} and \textbf{\tt movie-2.avi}:
Helical structure of the magnetic field streamlines around vortices in noncentrosymmetric 
superconductors. The fisrt movie (movie-1.avi) shows the helical structure of the streamlines 
for moderate value of the parity-breaking coupling $\gamma/k\lambda=0.2$. The streamlines 
here feature all the same chirality. The second animation (movie-2.avi) corresponds to 
rather important parity-breaking coupling $\gamma/k\lambda=0.8$, for which the longitudinal 
component of the magnetic field is inverted at some distance from the core. The chirality 
of the streamline depends on whether the longitudinal component of the magnetic field is 
inverted. That is depending on the chirality, some of the streamlines propagate forward 
(along positive $z$-direction), while other propagate backward (along negative $z$-
direction).

\item \textbf{\tt movie-3.avi}:
Magnetic streamlines, emphasizing forward propagating lines (solid tubes), in the case 
of field inversion due to important parity-breaking coupling $\gamma/k\lambda=0.8$. The 
transparent tubes propagate backward (along the negative $z$-direction).

\item \textbf{\tt movie-4.avi}:
Magnetic streamlines, emphasizing backward propagating lines (solid tubes), in the case 
of field inversion due to important parity-breaking coupling $\gamma/k\lambda=0.8$. The 
transparent tubes propagate forward (along the positive $z$-direction).

\end{itemize}

\section{Calculation of the integrals}
\label{Sec:Integrals}

The intervortex interaction \Eqref{Eq:Interaction}, the components of the magnetic field 
\Eqref{Eq:London:B:sol:0}, and the components of the current \Eqref{Eq:London:J:sol:0} 
are expressed in terms of integrals of the generic form 
\Equation{EqApp:Integral}{
G_\nu(x)\!=\! \int_0^\infty\! \frac{P(q)}{(1+q^2)^2-4\Gamma^2q^2}q^{\nu+1}J_\nu(qx)dq
\,,
}
where $\nu=0,1$. Introducing the complex number $\eta=\Gamma-i\sqrt{1-\Gamma^2}$, 
the quotient of the polynomials $P(q)$ and $(1+q^2)^2-4\Gamma^2q^2$ can be written 
as
\Equation{EqApp:Polynomial:1}{
\frac{P(q)}{(1+q^2)^2-4\Gamma^2q^2} = \frac{C}{q^2-\eta^2}+\frac{C^*}{q^2-\eta^{*2}} 
\,,
}
where $^*$ stands for the complex conjugation (since $q$ is real). The coefficient 
$C$ is the solution of the equation
\Equation{EqApp:Polynomial:2}{
P(q)=C(q^2-\eta^{*2})+C^*(q^2-\eta^{2}) \,.
}
The integral \Eqref{EqApp:Integral} becomes as follows:
\Align{EqApp:Integral:3}{
G_\nu(x)&= 2\int_0^\infty\!\Re\left[\frac{C}{q^2-\eta^2}\right]q^{\nu+1}J_\nu(qx)dq 
\nonumber\\
&= 2\Re\left[C\int_0^\infty\! \frac{q^{\nu}}{q^2-\eta^2}J_\nu(qx)qdq\right]
\,.
}
The integral in \Eqref{EqApp:Integral:3} has a form of an Hankel transform 
\cite{Piessens:18} which is an integral transformation whose kernel is a Bessel 
function. In short, the $\nu$-th Hankel transform $F_\nu$ of a given function $f(q)$ 
with $q>0$ is defined as 
\Equation{EqApp:HankelTR:0}{
F_\nu(x):=\int_0^\infty f(q)J_\nu(qx)q dq\,.
}
The inverse of the Hankel transform is also a Hankel transform. The functions
\Equation{EqApp:HankelTR:1}{
f(q)=\frac{q^\nu}{q^2+a^2} ~~~\longleftrightarrow~~~
F_\nu(x)=a^\nu K_\nu(ax)
\,,
}
are related to each other via the Hankel transformation~\eq{EqApp:HankelTR:0}.
Identifying $a=i\eta$, the integral in \Eqref{EqApp:Integral:3} thus read as
\Equation{EqApp:Integral:4}{
\int_0^\infty\! \frac{q^{\nu}}{q^2-\eta^2}J_\nu(qx)qdq=(i\eta)^\nu K_\nu(i\eta x)
\,.
}%
This is a well-defined expression since the constant $\eta$ is a complex number and 
the integral does not cross any pole. As a result, the integral \Eqref{EqApp:Integral} 
reads as follows:
\Equation{EqApp:Integral:5}{
G_\nu(x)=2\Re\left[C(i\eta)^\nu K_\nu(i\eta x) \right]
\,.
}
This generic relation determines both $\j$, the magnetic field $\B$ and the interaction 
$U(x)$. 
Notice that the modified Bessel function of the second kind in Eq.~\eq{EqApp:Integral:5} 
can be related to the Hankel function of first kind using the relation
\Equation{EqApp:Integral:6}{
K_\nu(z)=\frac{i}{2}\Exp{i\nu\pi/2}H_\nu^{(1)}(iz)\,,
~~\text{if}~-\pi<\mathrm{arg}\,z\leq\pi/2\,.
}
In our case,  $z=i\eta x$ and therefore $\mathrm{arg}\,z \in [0,\pi/2]$. 
Below we calculate the potential $U(x)$ as an example.

\subsection*{Example: calculation of the interaction}

The interaction $U(x)$, defined in the main body in equation \Eqref{Eq:FreeEnergy:K4}, 
reads as
\Align{}{
U(x)&=\int_0^\infty\! \frac{P(q)}{(1+q^2)^2-4\Gamma^2q^2} 
	\,J_0(q x)qdq \nonumber \\
\text{where}~~P(q)&=1+q^2. 
}
Solving \Eqref{EqApp:Polynomial:1} for the given polynomial $P(q)$, gives the coefficient 
$C=i\eta/(2\sqrt{1-\Gamma^2})$. The solution \Eqref{EqApp:Integral:5} for this 
particular problem is thus
\Equation{}{
U(x)=\Re\left[\frac{i\eta}{\sqrt{1-\Gamma^2}} K_0\left(i\eta x\right) \right]\,.
}
In the absence of parity-breaking $\Gamma=0$, $\eta=-i$. This provides the textbook 
expression for the interaction of vortices in the London limit: $U(x)=K_0(x)$. 
Similar calculations determine the solutions \Eqref{Eq:London:B:sol} and 
\Eqref{Eq:London:J:sol} for the components of $\B$ and $\j$.

\end{document}